\PassOptionsToPackage{dvipsnames}{xcolor}
\documentclass[nonacm,sigconf,10pt]{acmart}

\setcopyright{acmcopyright}

\acmDOI{10.475/1234.5678}

\acmISBN{123-4567-24-567/08/06}

\acmYear{2024}
\copyrightyear{2024}

\acmPrice{15.00}

\usepackage{tikz}
\usepackage{amsmath}
\usepackage{graphicx}
\usepackage{subcaption}  %
\usepackage{setspace}
\usepackage{soul}
\usepackage{enumitem}
\usepackage{titlesec}

\begin{document}

\addtolength{\floatsep}{-4mm}
\addtolength{\textfloatsep}{-3mm}
\addtolength{\dbltextfloatsep}{-10mm}
\titlespacing*{\section}{0pt}{1ex plus 0ex minus 0.5ex}{1ex plus 0ex minus 0.5ex}
\titlespacing*{\subsection}{0pt}{1ex plus 0ex minus 0.5ex}{1ex plus 0ex minus 0.5ex}
\titlespacing*{\subsubsection}{0pt}{1ex plus 0ex minus 0.5ex}{1ex plus 0ex minus 0.5ex}
\widowpenalty=0
\clubpenalty=0

\newcommand{\load}{\texttt{load}}
\newcommand{\store}{\texttt{store}}

\newcommand{\ignore}[1]{}

\newcommand{\myitem}[1]{\item \textbf{#1}}
\newcommand{\refalg}[1]{Algorithm~\ref{#1}}
\newcommand{\refeq}[1]{(Equation~\ref{#1})}
\newcommand{\reffig}[1]{Figure~\ref{#1}}
\newcommand{\reflst}[1]{Listing~\ref{#1}}
\newcommand{\refsec}[1]{Section~\ref{#1}}
\newcommand{\reftab}[1]{Table~\ref{#1}}
\newcommand{\refapp}[1]{Appendix~\ref{#1}}
\newcommand{\refln}[1]{Line~\ref{#1}}

\newcommand{\zhcom}[1]{{\sffamily\color{Orange}{ \scriptsize{}}}}
\newcommand{\smahar}[1]{{\sffamily\color{BrickRed}{ \scriptsize{}}}}

\title{CXL Shared Memory Programming:\\ Barely Distributed and Almost Persistent}

\author{Yi Xu}
\affiliation{%
  \institution{UC Berkeley}
}

\author{Suyash Mahar}
\affiliation{%
  \institution{UC San Diego}
}

\author{Ziheng Liu}
\affiliation{%
  \institution{UC San Diego}
}

\author{Mingyao Shen}
\affiliation{%
  \institution{UC San Diego}
}

\author{Steven Swanson}
\affiliation{%
  \institution{UC San Diego}
}

\renewcommand{\shortauthors}{Y. Xu et al.}

\begin{abstract}
While Compute Express Link (CXL) enables support for cache-coherent shared memory among multiple nodes, it also introduces new types of failures---processes can fail before data does, or data might fail before a process does. 
The lack of a failure model for CXL-based shared memory makes it challenging to understand and mitigate these failures.

To solve these challenges, in this paper, we describe a model categorizing and handling the CXL-based shared memory's failures: data and process failures.
Data failures in CXL-based shared memory render data inaccessible or inconsistent for a currently running application. We argue that such failures are unlike data failures in distributed storage systems and require CXL-specific handling. To address this, we look into traditional data-failure mitigation techniques like erasure coding and replication and propose new solutions to better handle data failures in CXL-based shared memory systems.
Next, we look into process failures and compare 
the failures and potential solutions with PMEM's failure model and programming solutions. We argue that although PMEM shares some of CXL's  characteristics, it does not fully address CXL's volatile nature and low access latencies. Finally, taking inspiration from PMEM programming solutions, we propose techniques to handle these new failures.

Thus, this paper is the first work to define the CXL-based shared memory failure model and propose tailored solutions that address challenges specific to CXL-based systems.
\end{abstract}

\maketitle

\sloppy

\section{Introduction}

In the last few years, Compute Express Link (CXL)~\cite{cxl} has emerged as a promising solution in datacenters for low-latency interconnect. CXL includes several new features like support for cache-coherent accesses, low-latency communication, and shared memory among multiple hosts and devices. One of the exciting new system architectures enabled by CXL is the ability to share memory among multiple hosts in a cache-coherent manner. CXL's memory sharing enables CXL-connected hosts to share and access a memory region while the hardware handles the coherency.

However, applications today are not designed to exploit this multi-host, cache-coherent shared memory. Current applications are typically either tailored for single-machine execution (e.g., microservices), where the applications are stateless and access globally shared data using well-defined APIs, or distributed in nature (e.g., databases) and use fault tolerance to maintain data availability and consistency.

To address these challenges, we categorize and define the CXL-based shared memory's failure model where data and processes can fail independently:
\begin{enumerate}[leftmargin=0.5cm]
    \item \textbf{Data Failures in CXL Systems.} Data failures occur when CXL memory becomes unavailable to other compute nodes. We argue that while data failures resemble failures in high-availability distributed systems, CXL-based shared memory has new, unique challenges.
    \item \textbf{Process Failures in CXL Systems.} Process failures occur when a process accessing CXL memory fails, but the data remains accessible. We argue that in a CXL system, if a process fails in the middle of an update to CXL-attached shared memory, it can result in inconsistent data. 
\end{enumerate}

We use this failure model to propose mechanisms to handle data and process failures. Though data failures in CXL-based multi-host data stores resemble distributed storage systems in how they enable storing and sharing of data, they differ dramatically in handling failures and redundancy. As we will discuss in \refsec{subsec:data-failure}, techniques like erasure coding and replication have limitations for CXL systems. For example, unlike traditional networked storage systems, CXL systems use \load{}/\store{} which introduce additional CPU overhead for replication compared to offloading these operations to the NIC's DMA engine. To address this, we propose a CXL switch-based replication mechanism.

Further, while process failures in CXL resemble the persistent memory (PMEM) failure model in certain aspects, PMEM’s solutions are not sufficient due to the differences in PMEM and
CXL’s failure models, performance, and persistence guarantees (\refsec{subsec:process-failure}). For example, using PMEM techniques like undo- or redo-logging while address data-consistency challenges, they add significant additional overhead with no persistency guarantees. To address the challenges of achieving data-consistency in the face of failures, we propose using whole-system persistence with minimal modifications to avoid the high overhead of logging.

Next in \refsec{subsec:why-both}, in addition to independently addressing process and data failures, we also propose specific techniques to mitigate them together to simplify the implementation and improve their efficiency. This is because process-failure mitigation techniques like logging and checkpointing maintain multiple copies of application data which could help lower the overhead of data-failure mitigation techniques. Finally in \refsec{subsec:free-side-effects}, we propose and discuss the added advantages of mitigating data and process failures for process migration and improved data access bandwidth.

\section{Failure Model}
\label{sec:failure-model}

In this section, we describe the failure model of CXL-based shared memory systems.

A CXL-based system introduces unique independent failures in nodes that contain only data or only processing logic,
because it enables physical separation of data and processing logic across nodes.
CXL enables practical memory disaggregation by allowing compute resources to use \load{}/\store{} instructions to access memory on other devices.
Thus, in a CXL-based disaggregated memory system, most memory is decoupled from compute resources~\cite{directCXL22}.

We describe the CXL failure model by dividing the system into multiple failure domains, each consists of devices susceptible to simultaneous failures from a mutual source.

\begin{figure}
\centering
\includegraphics[width=0.95\linewidth]{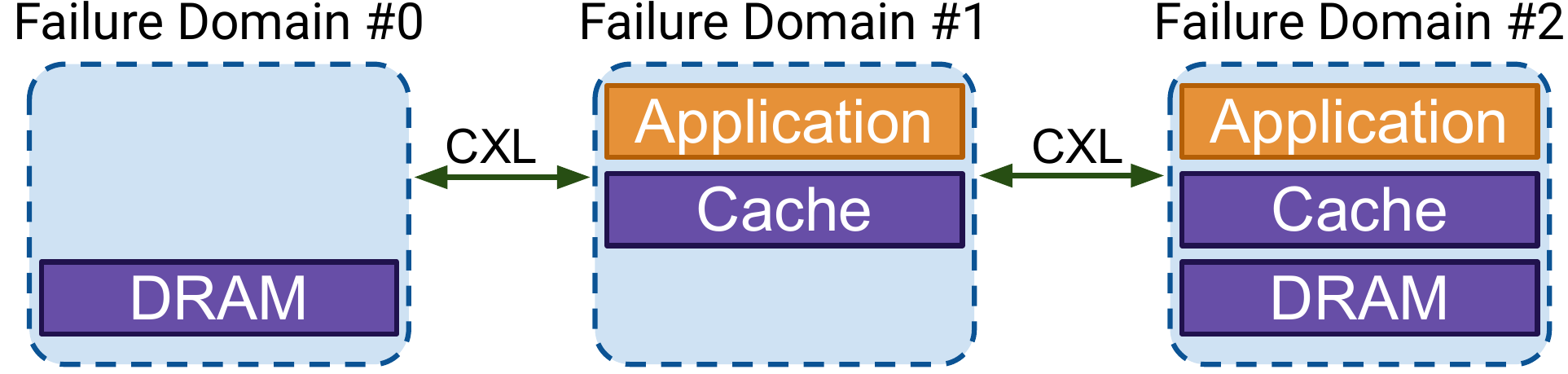}
\caption{\label{fig:cxlfailure} CXL's Failure Model. 
Three failure scenarios based on the application's location and data.
(1) Domain \#0 with data: data failure. (2) Domain \#1 with execution context: process failure. (3) Domain \#2 with data and execution context: data and process failure.}
\end{figure}

As Figure~\ref{fig:cxlfailure} shows, 
as application and data may or may not reside in 
the same failure domain,
the CXL failure model has three scenarios:
data failure (domain \#0), process failure (domain \#1),
and both data and process failure (domain \#2).

\textbf{Data failure} occurs when a node that contains data fails, rendering the data inaccessible to all CXL-connected processing devices. 
Further, data failure may also make data inconsistent, crucial for systems using CXL to communicate and share data with other connected hosts~\cite{zhang2023partial, mahar2024telepathic}. 
If a CXL system stores data on several devices
that are in different failure domains, 
the likelihood of one of the devices failing increases with the number of devices. 
Without fault tolerance mechanisms, applications utilizing CXL memory 
could experience significantly higher failure rates compared to 
those using only local memory.
Thus, a practical CXL memory system must provide a scalable and 
efficient fault tolerance mechanism in the event of unexpected data failures.

\textbf{Process failure} occurs when a node containing processing logic fails.
When a process fails, it might be in the midst of updating remote memory, leaving the data inconsistent. 
The CXL hardware ensures atomicity at the cache line granularity~\cite{zhang2023partial,patil2023apta}, 
but applications may define atomicity differently 
based on specific operation semantics. 
In the event of a failure, remote updates 
may remain partially completed, leading to potential 
inconsistencies in the data.

\begin{figure}[t]
\abovecaptionskip
\belowcaptionskip
\centering
\begin{subfigure}{0.25\columnwidth}
  \includegraphics[width=0.95\linewidth]{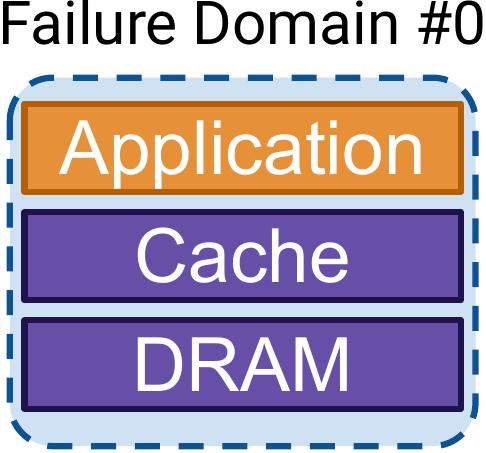}
  \caption{Single}
  \label{fig:singlemachine}
\end{subfigure}
\hfill
\rule{0.5pt}{1.5cm}
\hfill
\begin{subfigure}{0.62\columnwidth}
  \includegraphics[width=0.95\linewidth]{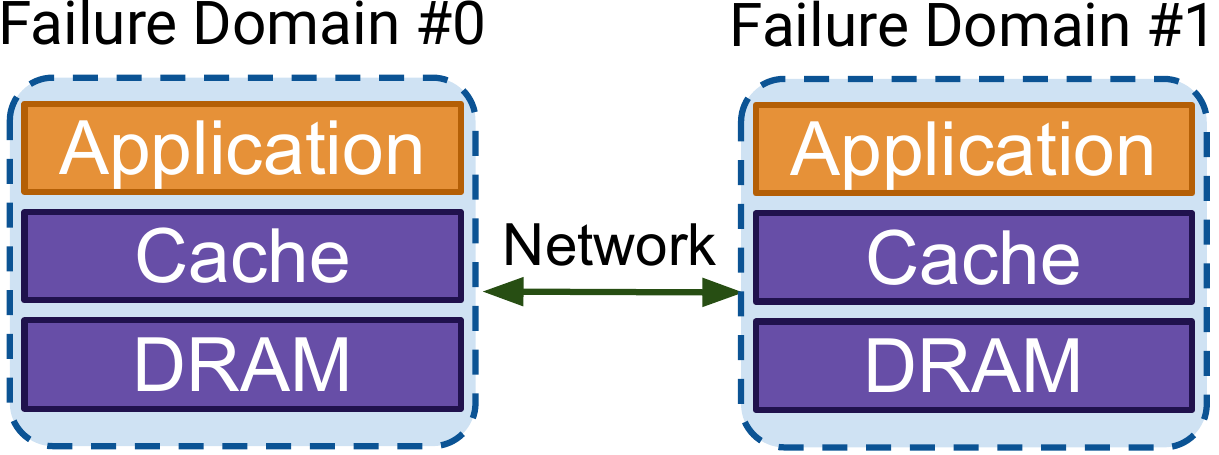}
  \caption{Distributed System}
  \label{fig:distributed}
\end{subfigure}
\hfill
\caption{Failure Models of Traditional Systems.}
\label{fig:existing}
\end{figure}

This failure model is different from those of most traditional systems.
As depicted in Figure~\ref{fig:singlemachine}, 
a single-machine system only has one failure domain, 
where the data and processing logic fail together.
Though distributed systems have multiple nodes that can fail independently---the data and processing logic usually coexist on the same node.
Thus, a failure still crashes both data and processing logic concurrently in typical distributed systems,
as shown in Figure~\ref{fig:distributed}.

Thus, CXL-based shared memory complicates failures and current mechanisms for managing failures in distributed systems do not align well with CXL. Existing mechanisms might not address both the data and process failures, 
could result in data inconsistency, or may overlook the potential to salvage data that is still consistent. Furthermore, CXL's high bandwidth and low latency nature presents new possibilities for rapid recovery and more efficient fault tolerance.

\section{Handling Failures in CXL-based System}
\label{sec:handling-failures}

In this section, we describe mechanisms to handle 
different types of failures. We demonstrate their 
infrastructure, discuss how they can be co-designed for data and process failures, and highlight some benefits that come with them. 

\subsection{Data Failure}
\label{subsec:data-failure}
Data failures in CXL-based systems could be managed similarly to how they are handled in distributed systems, but with adaptations to solve challenges introduced by CXL.

One of the most common ways of ensuring data availability in distributed systems is replication~\cite{xu2024ionia, liskov1991replication, leslie2001paxos, ganesan2021exploiting, ongaro2014search}, 
where multiple copies of data are stored across
nodes.

Besides known drawbacks of replication such as high storage overhead, new challenges arise within CXL environments.
As hosts access data in CXL-attached memory devices using \load{} and \store{} instructions, the large amount of data transmission of replication requires active CPU usage, unlike in distributed systems where operations
can be dispatched and queued in the NIC.

Replication in memory-based systems is also challenging because systems expect memory to have low latency.   
To meet this expectation, we propose embedding replication functionality in the CXL network by augmenting CXL switches~\cite{xconn2023hybrid} to replicate \store{} and steer \load{} to balance memory traffic.
A compute node issues stores to a CXL switch as it would to a shared memory node and the switch would replicate the stores across memory devices in multiple failure domains. To ensure even greater redundancy, the compute node could be configured to issue stores to multiple redundant CXL switches similar to high-availability storage systems~\cite{flexpod2023architecture}.
In this way, the CPU can dispatch \load{}/\store{} operations to the switch \ignore{ as it would for a normal CXL-attached memory device} and avoid the extra CPU usage introduced\ignore{by replicating directly from the compute node}.

Another widely used method is erasure coding, 
which involves breaking data into shards and generating parity shards~\cite{erasureCoding} that can be used to reconstruct the original data shards.
Compared to replication, erasure coding consumes less storage space but requires the data to be managed in fixed-size blocks.
Although data in CXL is transmitted at 256~B granularity (the FLIT size)~\cite{sharma2023introduction}, 
applications allocate memory objects of different sizes.
Thus, write operations using erasure coding would amplify the transmission from FLIT/object size to the whole block in order to compute the new parity block.
\ignore{which makes it an ill-fit for irregular memory management or accesses.}
Many memory-based erasure coding mechanisms solve this problem to some extent by organizing multiple objects into fixed-size groups, 
and then generating parity data for each group~\cite{cheng2019coupling, zhou2022carbink}.
These \textit{grouping strategies} developed for far-memory systems 
can be extended to CXL-based memory systems,
offering a potential solution.

However, erasure coding with grouping strategies may still need further improvements for CXL systems,
because it not only requires additional programming efforts and has performance overheads,
but also adds computational overhead 
 and is on the critical path. 
Due to the low access latency of CXL memory, 
the computational overhead now may constitute a 
significant portion of the overall overhead.

\subsection{Process Failure}
\label{subsec:process-failure}

We propose different process failure handling methods depending on the application characteristics. 
They address data inconsistency after process failures,
a significant concern in CXL-based shared memory systems for rapid process failure recovery~\cite{baumstark2021instant}.
This concern is also faced by PMEM systems, so we take inspiration from PMEM techniques.

However, directly applying PMEM techniques is not efficient, because CXL systems have different architectures and performance characteristics, unveiling different trade-offs, new challenges and new opportunities. 
Therefore, we re-examine and adapt PMEM techniques for CXL.

\subsubsection{Background of Data Consistency.}

The notion of failure atomicity is centered around preserving data consistency after failures
in both CXL and PMEM systems. 
It guarantees that a single operation 
or all operations in a transaction are either entirely applied or not applied at all,
thereby preventing data inconsistencies after failures.

In most PMEM systems, 
a critical aspect of ensuring data consistency
is maintaining \textit{write ordering}: writes need to be 
persisted in the correct sequence.
Therefore, these systems often integrate 
mechanisms to maintain an additional copy of consistent data, such as journaling, checkpoints, and transaction logs, often leading to \textit{write-amplification}.
Following unexpected failures, they will restore the data to a
known and consistent state, either
by rolling back or completing the partial updates. 
However, write ordering and write amplification lead to significant performance overhead in PMEM systems. 
Additionally, the need to identify transactions in every program demands substantial programming effort.

\subsubsection{Application-dependent Methods.}
For systems designed to execute programs that are open to program analysis and modification, and run workloads with known data access patterns, we propose using logging mechanisms.
Typically, applications can use three types of logs: undo~\cite{izraelevitz2016failure, scargall2020pmdk, go-pmem, atlas, mahar2023puddles}, redo~\cite{Mnemosyne,mahar2023puddles}, and resumption logs~\cite{izraelevitz2016failure, xu2021clobber, liu2018ido}. 

Unlike PMEM where the storage media is inherently slow,
the performance of logging in CXL systems depends on the location of log entries,
which is different between log techniques.
This shifts the balance of trade-offs between the three types of logs compared to PMEM.

\textbf{Read-heavy applications should choose undo log.} 
In CXL systems, the undo log must be located within a failure domain outside of the process node. 
Before each in-place update, the application must log the old value and ensure the log entry has arrived at the intended failure domain. 
\textit{Thus, undo log entries cannot be cached on the process node's CPU caches, which may have a substantial performance impact on write-heavy workloads.}
If the programs have data dependency inside transactions and are open to analysis, the amount of logging can be further optimized by \textit{resumption logs}.

\textbf{Write-heavy applications should choose redo log.} 
Redo log applications directly update log entries and only need to ensure entries have arrived in a different failure domain by the end of transaction. 
Then they can use logs to update the application data and commit the transaction. 
So there is no constraint preventing a log entry from being in the same failure domain until the end of transaction. 
\textit{Therefore, writes with redo logs are efficient since hot data can be cached locally, but reads may be slower because each read needs to check the updates in logs and might require redirection.}

\subsubsection{Application-independent Methods.}
For systems designed to execute programs that may not be open to analysis and modification, with unclear data access patterns,
we describe the checkpoint mechanism adapted from PMEM systems~\cite{elnawawy2017efficient, ren2015thynvm}, 
and advocate for a method inspired by whole system persistence techniques~\cite{narayanan2012whole,hodgkins2023zhuque}.

The CXL system could identify execution points at which data is consistent across threads 
and duplicate the data as checkpoints to a different failure domain. 
Updates to checkpoints must also be atomic.
Performance optimization 
can be achieved through incremental~\cite{Kim2022ListDBUO, wu2023treesls} or partial checkpoints. 
Unlike logging mechanisms that need to log most \store{} operations, 
the frequency of checkpoints can be adjusted to achieve lower overhead.
However, this method may double memory usage and, due to thread synchronization and high data volumes at checkpoints, can significantly increase tail latency.

Thus, we advocate a whole process persistence mechanism, which makes processes seamlessly continue 
execution as if the failure never occurred.
Whole process persistence persists only
the cache and registers.
There can be two implementations.
The hardware one requires a small backup power supply, 
and migrates the cache and register files to another failure domain when a process fails,
which has no performance cost before failure.
The software one maintains a reasonably up-to-date copy of the cache and registers in a different failure domain, with performance cost depending on the copy interval.

This approach allows 
more cost-effective recovery for CXL systems, with significantly smaller 
memory consumption compared to checkpoints. 
Compared to ordered cache eviction, persisting cache and registers in an unordered manner is simpler and faster.
It ensures IO consistency with the external world across failures: 
if an IO operation has been initiated, 
it cannot be undone or redone, 
as the system lacks the knowledge of whether it has been issued, 
along with its status and visibility to the external world.
Moreover, it provides building blocks to make failures 
transparent to users. 

\subsubsection{Key Insights.}
We share three insights about the metadata (e.g., the log, checkpoint, and cache and register dump) transferred by these methods.

The system should choose the metadata location by its design goal.
If designed for high availability of data, 
the metadata must be stored across failure domains by mechanisms in Section~\ref{subsec:data-failure}.
If designed for rapid recovery of processes, 
the metadata can be placed in the data node,
offering better performance and less memory consumption. 
After one processing node fails,
the metadata would rapidly restore the inconsistent data and other processing nodes can resume.

Designers should choose and tweak methods based on the metadata transfer cost, which may consider bandwidth, latency and data transmission fee.
For high transfer cost,
they should find logging more efficient,
and reduce the frequency of 
checkpoints and cache and register dumps.
For low transfer cost,
write amplification and programming complexity become more important, 
making logging less appealing.

Last, systems capable of handling data failures inherently gain some resilience 
to process failures. 
This insight enables co-design to effectively mitigate the overall performance 
overhead,
which we will expand in Section~\ref{subsec:why-both}.

\subsection{Handling Process and Data Failure \\Together}
\label{subsec:why-both}

In this section, we study how process and data failure handling interact with each other. Then we will look at several techniques for applications to integrate them together.
We analyze the performance of each mechanism and 
compare it with the baseline version that do NOT handle data or process failures. We define $R$ as the number of replicas (or the memory overhead of erasure coding).
Assuming data nodes share bandwidth through a CXL switch with the process node, replication reduces throughput by $\frac{1}{R}$. Direct connections with separate bandwidths might \textbf{not} affect throughput. This section uses the switch-based shared bandwidth assumption.

\subsubsection{Layered Process and Data Failure Handling}

\paragraph{Replication/Erasure Coding with Undo Logging.}  In scenarios involving writing to replicas or implementing erasure coding, rather than directly modifying values, applications can employ logging techniques like undo logging. During the recovery of data, the replica or parity shard will be rolled back based on the log. 
Because logging doubles each write and imposes memory ordering,
this mechanism results in 2$\times$ latency and $\frac{1}{2R}$ $\times$ throughput.

\paragraph{Replication/Erasure Coding with Atomic Pointer Updates.} Alternatively, an application can update a copy of a data structure, and atomically update any pointers to point to this new copy. 
Depending on the data structure implementation and the update pattern,
this mechanism may result in $1$ to $n\times$ latency, and $\frac{1}{nR}$ to $\frac{1}{R}$$\times$ throughput.
Performance may decrease when frequently updating small fields of objects, requiring entire object copies for minor changes. This approach also demands substantial programming effort to maintain atomic operations.

\paragraph{Replication/Erasure Coding with Whole Process Persistence.}
If the hardware can introduce a temporary additional failure domain   (e.g., by using a battery to back up cache and registers) in the event of a process failure,
we could integrate whole process persistence with any data failure handling mechanism. 
Since whole process persistence only transfers a small amount of data (local cache and registers) 
across failure domains at the moment of process failure, 
it incurs no runtime overhead. 
This mechanism ensures $1\times$ latency and maintains a $\frac{1}{R} \times$ throughput.

\subsubsection{Co-designed Process and Data Failure Handling}

\paragraph{Redo-compacted Replication.}
In redo log systems, 
updates are directly applied to logs. 
Subsequent accesses to logged addresses are also redirected to log entries until the log is digested, 
and updates are propagated to the original data structure. 
Therefore, we could reduce the updates to replicas by redirecting all subsequent accesses to redo logs.

Background compaction processes the redo log into replicas to prevent indefinite size growth and improves read performance by avoiding searches through the entire log. This process includes reading and then garbage collecting the log entries.
This mechanism leads to $1\times$ write latency, as updates are complete upon log entry, and the compaction is off the critical path. Read latency exceeds $1\times$. Throughput varies between $\frac{1}{2R}\times$ and $\frac{1}{R}\times$, influenced by update skewness; it approaches $\frac{1}{R}\times$ with frequent updates to a few addresses.

\paragraph{Stage-based Replication.}
To manage process failures during replication, one can intentionally enforce replicas to apply updates in a specific order. As long as the majority of replicas are not currently undergoing updates, the system remains capable of handling concurrent process and data failures. In the event of either process or data failure, the system can recover by rolling back or rolling forward to the states of the replicas not undergoing updates, which can act as logs.

The majority of replicas not undergoing updates can exist in either the before-update state or the after-update state. The decision to roll back or forward is based on how many replicas finished the operation, which also decides if the operation is marked as completed. 
Replicas advance through pre-execution, in-execution, and post-execution stages. 
Assuming $2n+1$ replicas in total, at most n replicas can be in the in-execution stage. 
Each operation is considered completed only when the majority of replicas finish an operation. 
Consequently, the latency may be up to $2\times$ higher, 
depending on replica setup. The throughput reduction is always $\frac{1}{R}\times$.

\paragraph{Log-structured Memory on Checkpoints.} 
A simple failure handling method is to replicate checkpoints across failure domains. Checkpoints enable recovery from a process failure and replicating them ensures data availability even if a failure domain goes down. However, with traditional checkpointing, the application would lose any new updates since the last checkpoint.

To address this, we propose to replicate the most recent checkpoint along with a log that records any new updates since the last checkpoint. This log resembles the \textit{semantic log} used in PMEM programming~\cite{memaripour2020pronto}. This ensures that in case the process and data nodes fail, the application can still reconstruct the most up-to-date state using the checkpoint and the semantic log. Whenever an application creates a new checkpoint, it replicates the checkpoint across failure domains and records the subsequent updates in a \textit{replicated} semantic log-structured memory.
This method offers a latency of $1\times$. It can achieve a $\frac{1}{R}\times$ throughput 
if checkpoint updates are efficiently implemented.

\subsection{Additional Benefits of Handling Failures}
\label{subsec:free-side-effects}
Failure handling usually relies on relatively up-to-date data copies,
which may provide benefits such as faster process migration~\cite{milojivcic2000process} and improved data-access bandwidth.

For example,  a CXL-based replicated memory system can significantly simplify migrating processes across machines in a data center.
An orchestrator can kill the process and use its checkpoint or whole process persistence state to resume it on another machine. While to the outside world, the application has migrated to a new machine, to the application it seems like there was a simple process failure.

Further, data failure handling mechanisms like data replication and erasure coding also offer improved data access bandwidths. This is because data reads can be served from any one of the replicas in a replicated CXL-based memory system. Similarly, erasure-coded memory systems enable better write throughput for large writes than a single memory system as a single large write operation is striped as smaller writes across multiple CXL-attached memory systems.

\section{Conclusion}
\label{sec:conclusion}
CXL-based shared memory systems have data and processes in separate failure domains and fail independently of each other. In this paper, we have categorized CXL's failures into data failures and process failures.

We propose mechanisms to handle data failure, inspired by traditional distributed systems while adapted to the unique characteristics of CXL systems. Similarly, we looked into the data consistency challenges of process failures and how they relate to persistent memory, and proposed solutions that address the non-persistent nature of CXL-based shared memory.
Finally, we addressed the challenges of handling process and data failures together, and highlighted the additional benefits of failure handling mechanisms.

\bibliographystyle{unsrt}
\bibliography{paper}
\end{document}